\documentclass[12pt]{article}

\usepackage{graphicx}

\newcommand{\be}{\begin{equation}}
\newcommand{\ee}{\end{equation}}

\newcommand{\sumij}{\sum_{\langle i,j \rangle}}
\newcommand{\Jij}{J_{ij}}

\newcommand{\figref}[1]{\ref{#1}}
\newcommand{\eref}[1]{(\ref{#1})}

\newcommand{\mt}[1]{\langle #1 \rangle}
\newcommand{\md}[1]{\overline{#1}}

\begin{document}   
\title {
Numerical Simulations of the $4D$ Edwards-Anderson
Spin Glass with Binary Couplings}

\author{
Enzo Marinari\thanks{e-mail address:Enzo.Marinari@ca.infn.it}\,
and
Francesco Zuliani\thanks{e-mail address: Francesco.Zuliani@ca.infn.it}\\
\small
Dipartimento di Fisica and INFN, 
Universit\`a di Cagliari  \\
\small
Cittadella Universitaria,
S. P. Monserrato-Sestu Km. 0.7  \\
\small
09042 Monserrato (CA), Italy
}

\date{April 1999}

\maketitle
 
\begin{abstract}
\noindent We present numerical results that allow a precise
determination of the transition point and of the critical exponents of
the $4D$ Edwards-Anderson Spin Glass with binary quenched random
couplings.  We show that the low $T$ phase undergoes Replica Symmetry
Breaking.  We obtain results on large lattices, up to a volume
$V=10^4$: we use finite size scaling to show the relevance of our
results in the infinite volume limit.
\end{abstract}

\section{Introduction}

The question of whether short range Edwards Anderson (EA) spin glasses
share the remarkable features of the infinite range Sherrington
Kirkpatrick (SK) model \cite{SPINGLASS} is still an open one.  In this
work we present Monte Carlo simulations of the $4D$ EA Ising Spin
Glass \cite{BHAYOU,REBHYO,PARRIT,CIPARI,BCPRPR,PARIRU,BERCAM} with a bimodal
distribution of the quenched couplings, performed on large lattice
volumes (thanks to the tempering and parallel tempering simulation
technique \cite{PARTEM,OPTI}), with a large number of samples, and
down to low values of the temperature $T$.  In this way we are able to
obtain detailed information about the nature of the transition, to
determine with good precision critical temperature and exponents, and
to give strong evidence supporting the fact that the low-temperature
phase is mean-field-like. A great deal of effort has gone in ensuring
reliability of the data on delicate issues such as thermalization
checks and consistency of data analysis.

The paper is organized as follows: first of all we describe the model
and the parameters of our MC simulation.  We then present data related
to the Binder cumulant and to the determination of $T_c$ and $\nu$.
By analyzing the overlap susceptibility we determine the value of
$\eta$. Finally we discuss in detail about the probability
distribution of the overlap $P(q)$.  We present, among others,
evidence for non-triviality of single sample $P_J(q)$ and for a
non-zero value of the position of the maximum of $P(q)$, $q_{max}$, in
the thermodynamic limit.

\section{The Numerical Simulation}

It is very difficult to run reliable numerical simulation of finite
dimensional spin glasses.  The main reason for such difficulties is
the presence of many meta-stable states (responsible for aging effects
as well as for many other peculiarities of spin glasses
\cite{SPINGLASS}).  The Monte Carlo dynamics gets easily trapped, and
the system only probes a restricted part of phase space.

Many algorithmic solutions have been proposed to improve the speed of
thermalization of these systems.  All these techniques are related to
density scaling methods (see \cite{OPTI} for a review and references):
we use here the maybe simplest implementation of these ideas, the
parallel tempering~\cite{PARTEM}, where a number of configurations of
the system are allowed to exchange their temperature (for
multi-canonical methods, that are strongly related and have in
principle an even wider range of applicability, see for example
\cite{BERG}).  Thanks to parallel tempering we have been able to
thermalize systems of volume $V=10^4$ down to $T\simeq 1.2 (0.6 T_c)$.

We study the $4D$ Edwards-Anderson Ising spin glass with binary
couplings, with Hamiltonian

\begin{equation}
  H \equiv - \sumij \Jij \sigma_i \sigma_j \ ,
\end{equation}
where the sum runs over nearest neighboring sites, the $\sigma_i$ are
$\pm 1$ Ising spins, and the couplings are quenched variables drawn
with probability $\frac12$ among the two values 
$\{ -1, +1 \}$. 
The {\em overlap} among two different systems is defined as

\begin{equation}
  q^{\alpha, \beta} \equiv \frac{1}{V} \sum_i \sigma_i^{\alpha}
\sigma_i^{\beta} \ ,
\end{equation}
where $\alpha$ and $\beta$ denote two configurations of the system in
the same realization of the quenched disorder.
The overlap probability distribution for a given sample is

\begin{equation}
  P_J(q) \equiv \mt{\delta(q-q^{\alpha, \beta})} \ ,
\end{equation}
where $\mt{...}$ denotes the usual Gibbs average.  
Its average over samples is

\begin{equation}
  P(q) \equiv \md{P_J(q)} \ ,
\end{equation}
and its moments are defined as

\begin{equation}
  q^{(n)} = \md{\mt{q^n}} = \int dq\  q^n\  P(q)   \ .
\end{equation}
We always denote by $\mt{\cdots}$ the thermal averages
and by $\md{\cdots}$ the disorder averages.

Our simulation have been performed on a set of workstations, using a
multi-spin-coding program that was inspired by the work of
\cite{RIEGER}. We have selected the parameters of our Monte Carlo and
parallel tempering runs such to guarantee a complete thermalization of
the measured observables. We will discuss this issue in some detail.

Table (\ref{T-PARAME}) summarizes the relevant parameters used in the
simulation: we give among others the number of thermalization steps,
of measurements steps, the number of different disorder realizations
and the temperature ranges investigated by tempering.  Temperature
values have been chosen uniformly spaced in the interval between
$T_{min}$ and $T_{max}$.

\begin{table}
\centering
\vspace{3mm}
\begin{center}
\begin{tabular}{|c||c|c|c|c|c|c|c|} \hline
$L$ & Thermalization & Equilibrium & Samples & $N_{\beta}$ & $\delta T$ &  
$T_{min}$& $T_{max}$\\ \hline \hline
3    & 100000  &  100000  & 3200   &  17  &  0.1   & 1.2 & 2.8\\ \hline
4    & 100000  &  100000  & 2944   &  17  &  0.1   & 1.2 & 2.8\\ \hline
5    & 100000  &  100000  & 1920   &  17  &  0.1   & 1.2 & 2.8\\ \hline
6    & 100000  &  100000  & 1120   &  33  &  0.05  & 1.2 & 2.8\\ \hline
8    & 100000  &  100000  & 1376   &  33  &  0.05  & 1.2 & 2.8\\ \hline
10   & 150000  &  150000  & 512    &  56  &  0.04  & 1.2 & 3.4\\ \hline
\end{tabular}
\end{center}
\caption[0]{Parameters of the Tempered Monte Carlo runs.
\protect\label{T-PARAME}}
\end{table}

We have used different methods to verify that we have 
correctly thermalized the systems.  Using ``parallel tempering'', one is
actually performing a generalized Markov chain where systems at
different temperatures are allowed to ``move'' in temperature-space
too. A necessary condition for the Markov chain to be effective in
de-correlating different measurements is the fact that each system
spans at least a few times all the allowed temperature range during
the simulation. In this respect we check {\em a posteriori} that the
probability of swapping temperature has been of order $0.5$ (ensuring
in this way that a single system did not get stuck at a specific value
of $T$) and that the histogram counting the time that each system has
spent at each temperature is fairly flat.  This requirement is
fulfilled in all our simulations, for all $T$ and $L$ values.

Another very strong check of thermalization is the fact that the
single sample $P_J(q)$ are symmetric in the limits of the statistical
significance of the histogram. This is very well verified as can be
seen for example in figure \figref{fig:pjq} where we plot $P_J(q)$ 
for selected samples.

\section{The Binder Parameter, $T_c$ and $\nu$}

We start by discussing the overlap Binder parameter. We will use it to
qualify the phase transition, and to determine the critical
temperature and the first of the critical exponents, $\nu$. 
We will use and describe different methods to compute the quantities
we are interested in. Our statistical sample of configurations is a
large sample, and our set of data precise (even as far as the
dependence over the lattice volume $V$ is concerned): we will show
that different analysis styles give compatible (precise) results.

\begin{figure}
\centering
	\includegraphics[width=0.8\textwidth]{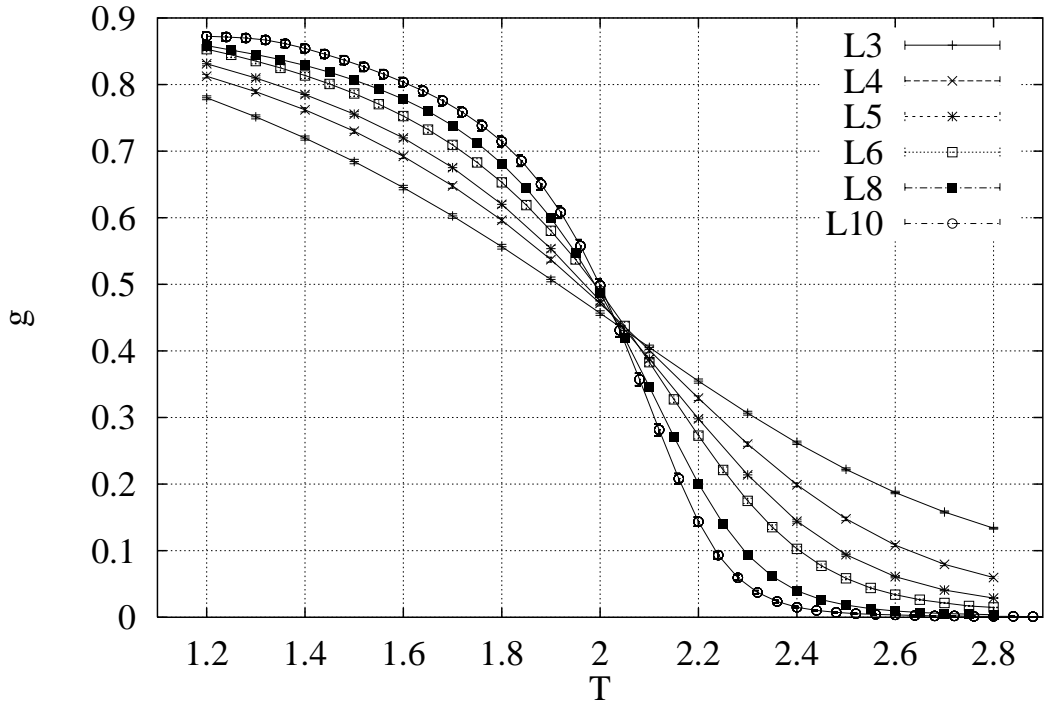}
\caption[qqq]{Binder parameter $g$ versus $T$, for different values of
the linear lattice size $L$ (see caption in the plot).}
\label{fig:binder_allsides}
\end{figure}

We define the usual overlap Binder parameter as

\begin{equation}
  g = \frac{1}{2} 
  \left( 3 - \frac{ \md{\mt{q^4}}}{\md{\mt{q^2}}^2}\right) \ .
\end{equation}
The Binder parameter is an adimensional quantity, and its value at the
critical point is universal. Close to $T_c$ its leading behavior is

\begin{equation}
  g(L,T) \simeq \bar{g}\left(L^{\frac{1}{\nu}}
  \left(T - T_c\right)\right) \ .
\label{eq:binderscaling}
\end{equation} 
In usual ferromagnetic systems the infinite volume limit of the
magnetization Binder cumulant is $0$ in the warm phase (where the
distribution of the order parameter is Gaussian) and $1$ in the broken
phase: for a spin glass with replica symmetry breaking and hence a
non-trivial distribution of the overlap order parameter, the
transition is signaled by a non-trivial value of $g$ in the broken
phase (in the warm phase one expects an infinite volume limit of
zero).  In both cases the location of $T_c$ is signaled by the
crossing of the curves of $g$ versus $T$ for different values of the
lattice size $L$ (asymptotically for large $L$): large $L$ curves are
lower for $T>T_c$ and higher for $T<T_c$. We show in figure
\figref{fig:binder_allsides} $g$ versus $T$ for different $L$ values.
The crossing point is close to $T\simeq 2$ for all lattice values, and
the value of the Binder cumulant at criticality, $g_c$ is close to
$0.45$. Also error analysis has been a sensitive issues. We have
always used a jackknife or a bootstrap error analysis \cite{JACKKNIFE}
{\em directly} on the fitted parameters to determine errors. Still one
has to keep in mind that statistical errors come together with
systematic errors, due to the functional form one decides to try to
fit (typically the asymptotic scaling form, that on finite size
lattices is affected by power corrections). The two types of errors
have to be kept under control separately.

In figure \figref{fig:binder_allsides} the crossing of the different
$g_L$ curves is very clear.  It is interesting to stress the
difference with the three dimensional case \cite{MAPARU} , where the
crossing at $T_c$ looks more like a merging of the different
curves. $3d$ is (very) close to the lower critical dimension, while in
$4d$ we are in a safe region: potentially this is important to make
the physical picture easier to understand.

Let us discuss a first naive approach to the data. By looking at the
crossing of the curves $g_L(T)$ versus $T$ for different $(L,L+1)$
values one sees that one cannot extract a systematic dependence of the
crossing point (and hence of the estimate of the effective critical
temperature $T_c(L,L+1)$) over $L$. Any systematic trend is smaller
than the statistical error (maybe just showing a systematic average
decrease of the estimate $T_c(L,L+1)$ when going from smaller to
larger $L$ values). The preferred value of $T_c$ is slightly larger
than $2.00$. A first naive estimate of $\nu$ can be done by
linearizing $g_L(T)$ around the estimate we have given for $T_c$, and
by evaluating the logarithm of the slope ratio (divided by the
logarithm of the two lattice sizes ratio,
$\log\left(\frac{L}{L+1}\right))$.  With this method, one gets a first
estimate for a set of effective exponents $\nu(L,L+1)$. Here too, one
cannot distinguish any clear strong dependence over the lattice size:
the error one gets on $\nu$ is completely correlated to the variation
of the estimate of $T_c$. For larger $T_c$ one estimates a lower value
of $\nu$, while for lower estimates of $T_c$ one gets larger estimates
for $\nu$. The error is dominated by this effect. The estimate for
$\nu$ is close to $1$.

To get a reliable estimate of $T_c$ and of $\nu$ we have used two
methods of analysis of $g$ (see for example the discussion of the
analysis of reference \cite{INPARU,MAPARU}).  In the first approach we
linearize the data close to $T_c$ (for all $L$ values) and we run a
global fit to all data: we fit $T_c$ and $\nu$ for the two variable
function $g_L(T)$ (as we said, linearized close to $T_c$). We use data
in a $T$ range around the interval $1.9-2.1$. We estimate the errors
over the fit parameters ($T_c$ and $\nu$) by a jackknife
approach~\cite{JACKKNIFE}: we repeat the fit approximately $K$ times
over a subsample of the data containing all of our statistical sample
but a fraction $\frac{1}{K}$. The error is estimated by looking at
fluctuations of the results of the $K$ fits, and by accounting for the
fact they are correlated \cite{JACKKNIFE}. We also repeat the fits by
discarding the smaller $L$ values, to check if we can observe any
systematic drift (again with good accuracy the average value of the
result does not seem to depend systematically over the $L$ range
selected).  Results are very stable, and the value we estimate for
$\nu$ systematically comes out to be close to $1.10$.

In the second approach, that comes in different flavors, one only uses
data in the warm phase. This method leads to a smaller statistical
error, that is balanced from a larger systematic incertitude (since
we only select data at a given distance from $T_c$, and approaching
$T_c$ leads to a systematic drift of the estimate). In this case we
start by selecting a threshold value for $g$, $g^* \le g_c$. We start
with low values of $g^*$, and we approach $g_c$ from below: we cannot
get too close to $g_c$ or the merging of the curves for different
sizes makes the error over the measurement too large (we use values of
$g^*$ going from $=0.2$ to $0.4$. We use a polynomial fit to
interpolate the data for $g(T)$, at different $L$ values. We have
decided to use a polynomial of degrees four (we have checked it
guarantees stable fits and consistent results), and we fit a $T$ range
in the critical region (for $L=3$ we use the data in the $T$ range
$1.5-2.8$, for $L=10$ we use the range $1.88-2.16$). We define now
$T_c(L,g^*)$ as the crossing point of the fitted polynomial with the
horizontal line at $g^*$, and $\nu^*$ as 

\begin{equation}
\lim_{L\to\infty} T_c(L,g^*)
= T_c(g^*) + \frac{A}{L^{\frac{1}{\nu^*}}}\ .
\end{equation}
When $g^*\to g_c$ $\nu^*\to \nu$. If $g^*$ is too small violations of
scaling are dominant, while if one approaches too much $g_c$ the
merging of the $g$ curves makes the error over the determination of
$\nu^*$ overwhelming. The errors have been estimated by using a {\em
bootstrap} approach (very similar in spirit to the jackknife
technique, see \cite{JACKKNIFE}): one emulates fake sets of data with
a Gaussian distribution around the real measurements, fits these
multiple sets of fake data and compute the errors over the fit
parameter.

We note at last that we have also used a variation of this second
method, described in \cite{MAPARU}, based on the direct analysis of
the derivative of $g$ with respect to $T$. Also this method gives
results that are compatible with the other ones.

\begin{figure}
\centering
	\includegraphics[width=0.8\textwidth]{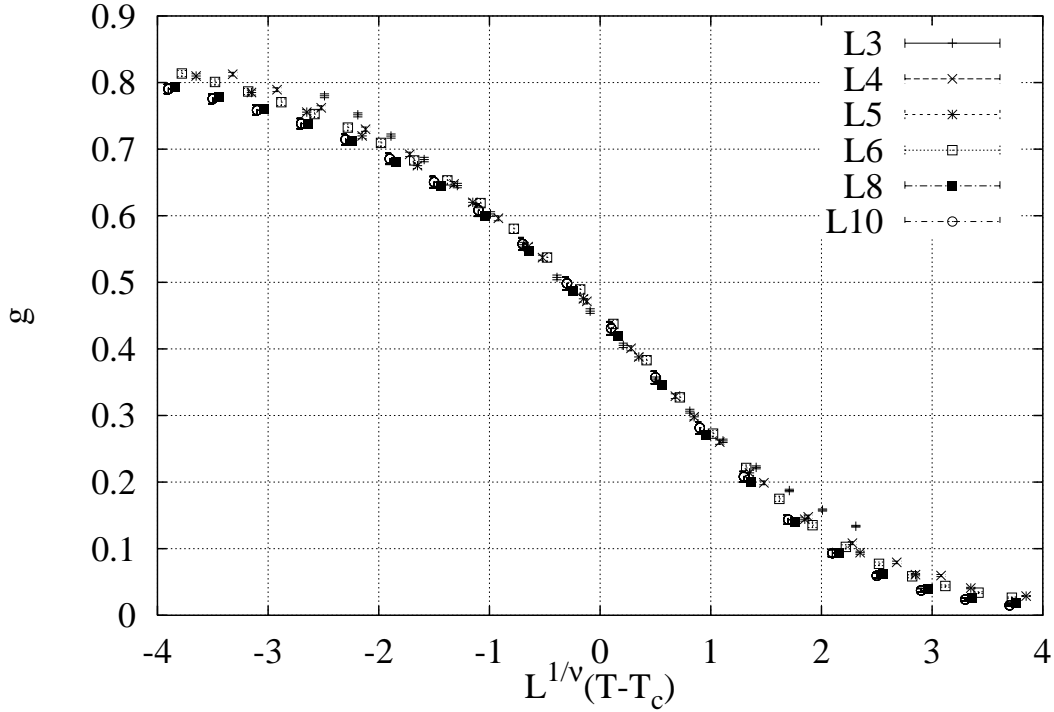}
\caption[qqq]{$g$ versus $L^{\frac{1}{\nu}}(T-T_c)$,
with $\nu=1.0$, $T_c=2.03$.}
\label{fig:binderrescaled}
\end{figure}

Our final estimates, averaged over the results obtained using these
different approaches, are

\begin{equation}
  T_c = 2.03 \pm 0.03\ ,
\end{equation}
and

\begin{equation}
  \nu = 1.00 \pm 0.10\ .
\end{equation}
In the rest of this paper we will use these two values as our best
estimates of $T_c$ and $\nu$.  We show in figure
\figref{fig:binderrescaled} the data for $g_L(T)$ rescaled by using
these two values: the scaling turns out to be very satisfactory.

\section{The Overlap Susceptibility and $\eta$}

\begin{figure}
\centering
\includegraphics[width=0.8\textwidth]{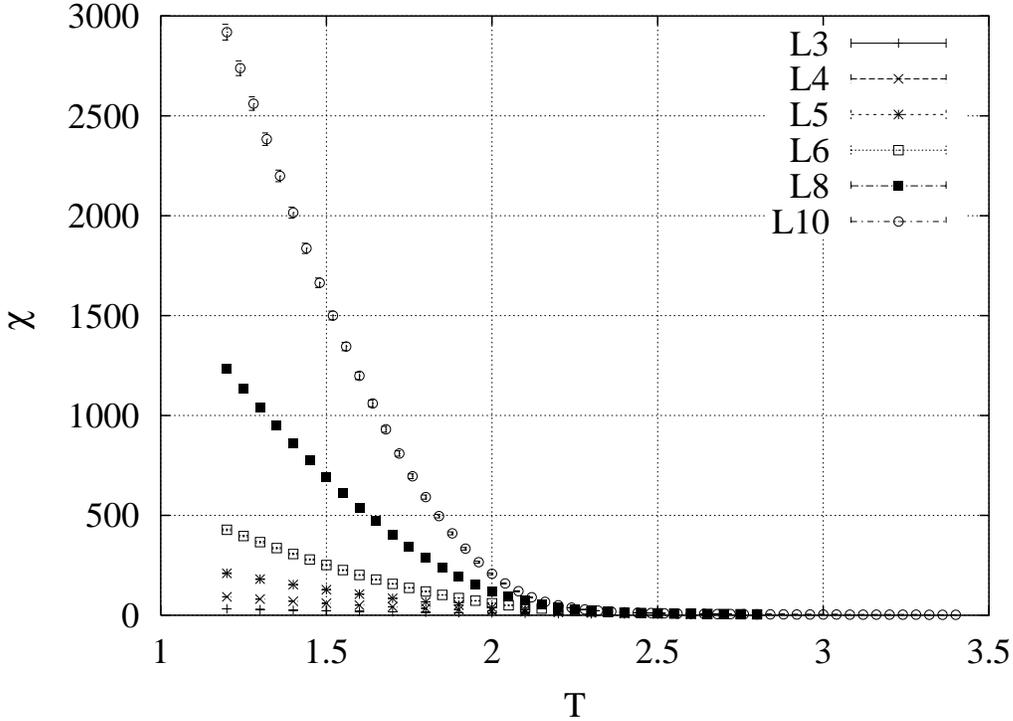}
\caption[qqq]{Overlap susceptibility 
$\chi_q$ versus $T$, for different $L$ values.}
\label{fig:susc}
\end{figure}

The determination of the overlap susceptibility, $\chi_q$, provides
various possible ways to determine the exponent $\eta$ (and hence of
the exponent $\gamma$). In a spin glass in the RSB phase the overlap
susceptibility

\begin{equation}
  \chi_q \equiv V \langle q^2 \rangle
\end{equation}
is expected to diverge for all values of $T\le T_c$. 
We show $\chi_q$ versus $T$ in figure \figref{fig:susc}.

The first method is
based on the fact that we expect that at $T=T_c$

\begin{equation}
  \chi_q(L,T=T_c) \simeq L^{2-\eta}\ .
\end{equation}
We use a linear interpolation of the data in the region close to
$T_c$. As in the case of $g$ the error in the estimate is mainly
related to the choice of $T_c$. The fit at $T=2.03$ by using $L>3$
gives an estimate of $0.28$.

\begin{figure}
\centering
\includegraphics[width=0.8\textwidth]{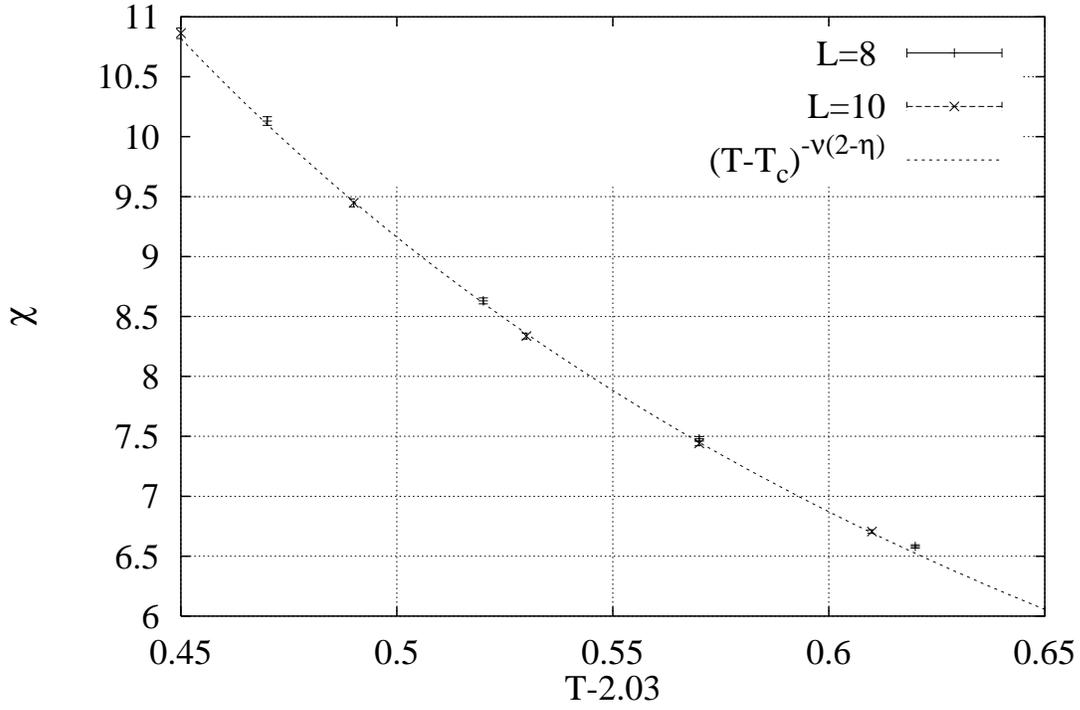}
\caption[qqq]{Best fit to the overlap susceptibility 
$\chi_q$ versus $T-T_c$, at $T>T_c$.}
\label{fig:susc_warm}
\end{figure}

In the second method we use data where $L\gg\xi$. We go as close to
$T_c$ as possible, under the condition that data on our larger lattice
($L=10$) coincide, in our statistical accuracy, with the ones at
$L=8$. Here we expect that

\begin{equation}
  \chi_q(T) \simeq (T-T_c)^{-(2-\eta)\nu}\ .
\end{equation}
We can use data down to $T=2.5$ (i.e. at a $\Delta T\simeq 0.5$ from
$T_c$), where finite size effect start to be sizable even at
$L=10$. We show our best fit (in a $T$ interval of $=.2$) in figure
\figref{fig:susc_warm}. In this region we have a stable fit, with
$\eta$ close to $-0.4$. Even if this second measurement is not very
precise (we have to stay quite far from the critical region) it is
interesting the fact that we get a coherent determination of $\eta$,
by using a completely different scaling region than in the former
analysis (the new analysis also depends on the value of $\nu$ we have
determined by using $g$).

The last approach we use for determining $\eta$ is based on the
analysis of the scaling properties of the distribution probability
$P(q)$ of the overlap order parameter $q$ in the region $q\simeq 0$ at
$T=T_c$. We analyze the behavior of $P(q)$ in the next section, but we
discuss now the scaling of $P(0)$ of $T_c$ in order to define our
determination of $\eta$. At $T=T_c$ we expect

\begin{equation}
  P(q\simeq 0) \simeq L^{\frac{d-2+\eta}{2}}\ ,
\end{equation}
i.e. in $d=4$ a scaling with $L^{\frac{2+\eta}{2}}$.
We find a very good best fit (we do not include the $L=3$ data), with
an $\eta$ value close to $-0.3$.

\begin{figure}
\centering
\includegraphics[width=0.8\textwidth]{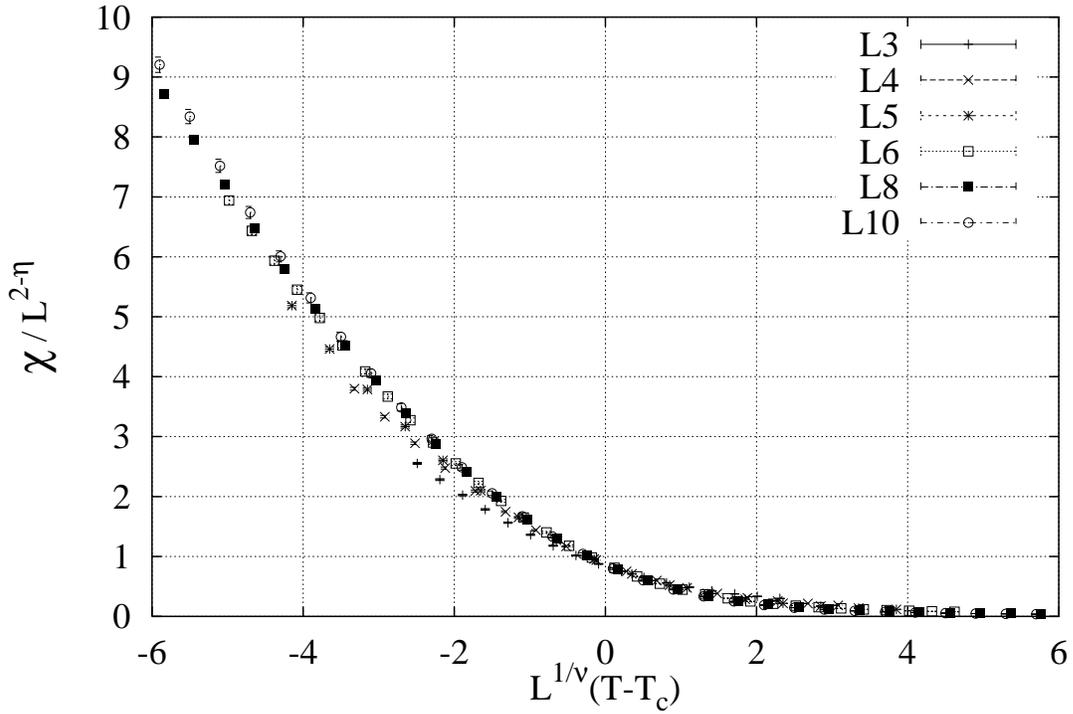}
\caption[qqq]{
  Rescaled overlap susceptibility $\frac{\chi_{SG}}{L^{2-\eta}}$ versus 
  $L^{\frac{1}{\nu}}(T-T_c)$, with $T_c=2.03$, $\nu=1.0$ and
  $\eta=-0.30$.}
\label{fig:suscrescaled}
\end{figure}

By considering all the methods we have discussed in this section we
give our final estimate

\begin{equation}
  \eta = -0.30 \pm 0.05\ ,
\end{equation}
that we will use in the rest of our analysis. In figure
\figref{fig:suscrescaled} we plot $\chi_q$ rescaled by using our best
fits. The rescaling works fine.

Let us also notice that we have a good agreement with the results
reported in \cite{PARIRU} for the $4d$ EA model with Gaussian
couplings. There the authors find $\nu\simeq 1.06$, and $\eta \simeq
-0.35$. Universality seems to work.

\section{$P(q)$}

\begin{figure}
\centering
	\includegraphics[width=0.8\textwidth]{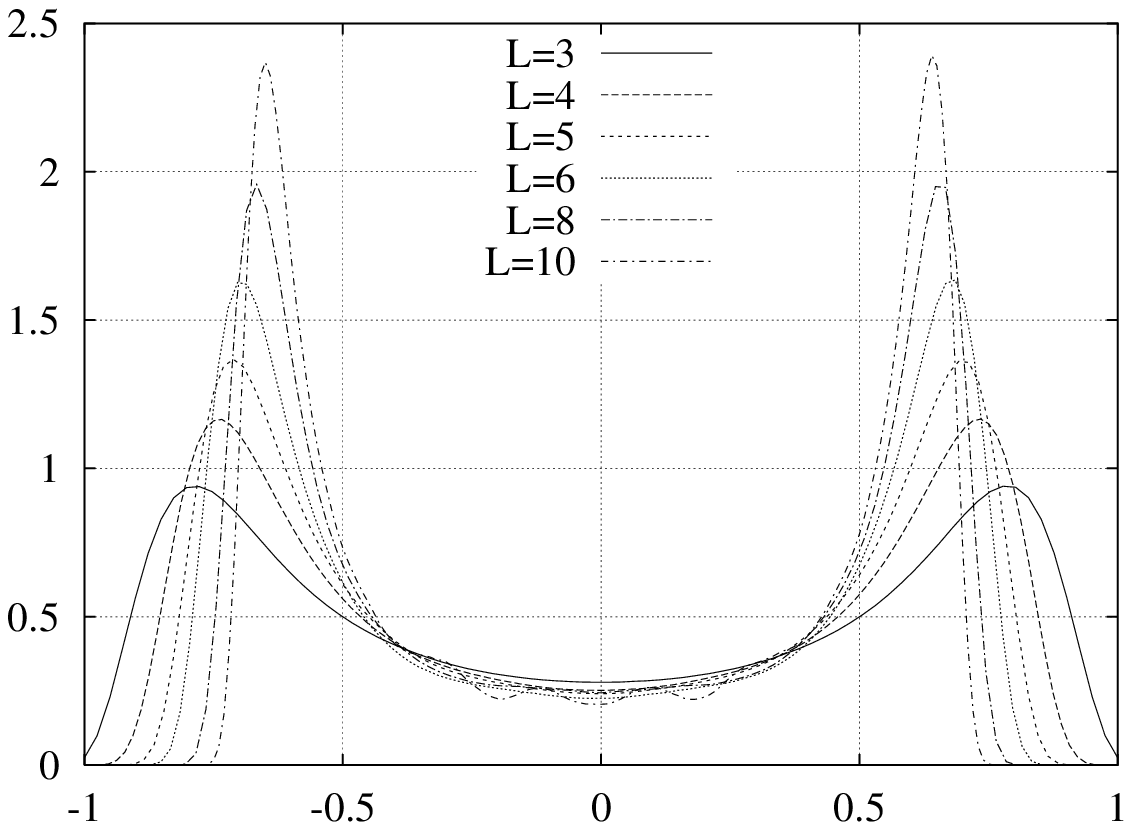}
\caption[qqq]{$P(q)$ at $T= 1.2$ (broken phase), 
for different lattice sizes.}
\label{fig:pqt12}
\end{figure}

In the two former sections we have shown that the $4D$ EA model
undergoes a phase transition, and we have determined its location and
the critical exponents. Now we will try to qualify it in better
detail, by determining and analyzing the probability distribution of
the order parameter, $P(q)$.

In figure \figref{fig:pqt12} we show our average $P(q)$ (averaged over
the different disorder realizations) at $T=1.2<T_c$. When increasing
the lattice size the peak where $P(q)$ is maximum shifts to lower $q$
values: for showing that there is a phase transition to a phase with a
non zero expectation value of $q$ we have to show that the peak does
not go to $q=0$ when $L\to\infty$. The {\em plateau} of $P(q)$ for
$q\simeq 0$ does not lower when increasing $L$, as we will discuss
better in the following. We remind the reader that in the RSB Parisi
Mean Field scenario the $P(q)$ is (in zero magnetic field) a non
trivial function, that in the infinite volume limit is formed by a
$\delta$ function at $q=q_{EA}$ and by a regular part that extends
down to $q=0$. On the contrary if the broken phase has the same
structure of the one of an ordered ferromagnet $P(q)$ has to become
asymptotically the sum of two $\delta$ functions at $\pm q_{EA}$.

\begin{figure}
\centering
	\includegraphics[width=0.8\textwidth]{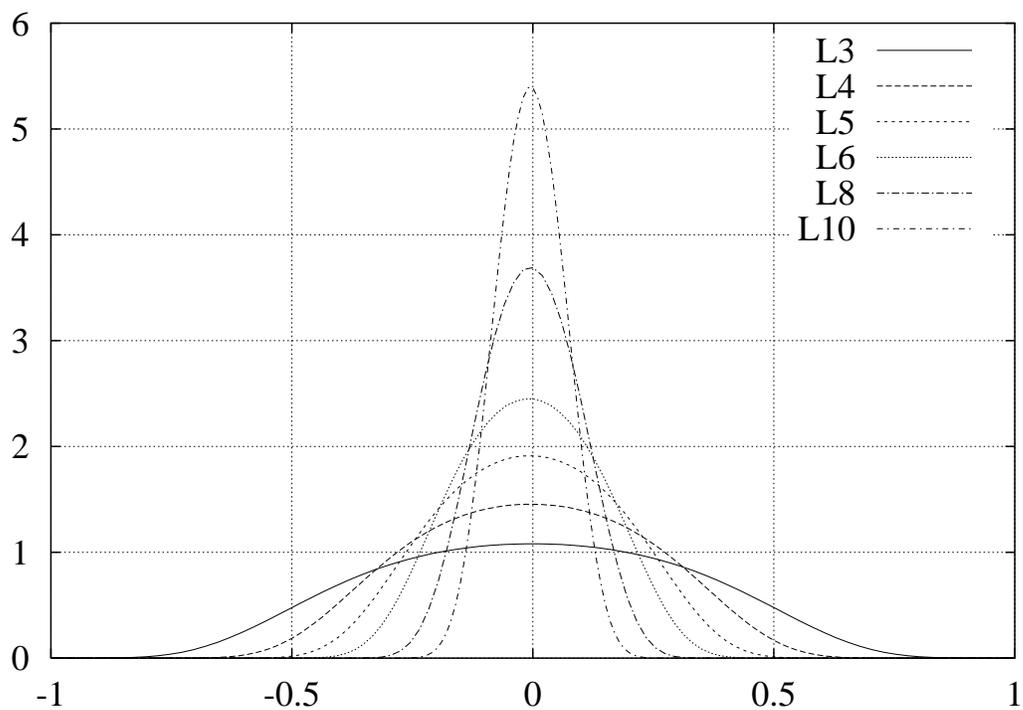}
\caption[qqq]{$P(q)$ at $T= 2.2$ (warm phase), 
for different lattice sizes.}
\label{fig:pqt22}
\end{figure}

For sake of comparison  we show in figure \figref{fig:pqt22} what
happens in the warm phase, where the average $P(q)$ shrinks to a
Gaussian distribution around $q=0$ when $L\to\infty$. 

\begin{figure}
\centering
	\includegraphics[width=0.8\textwidth]{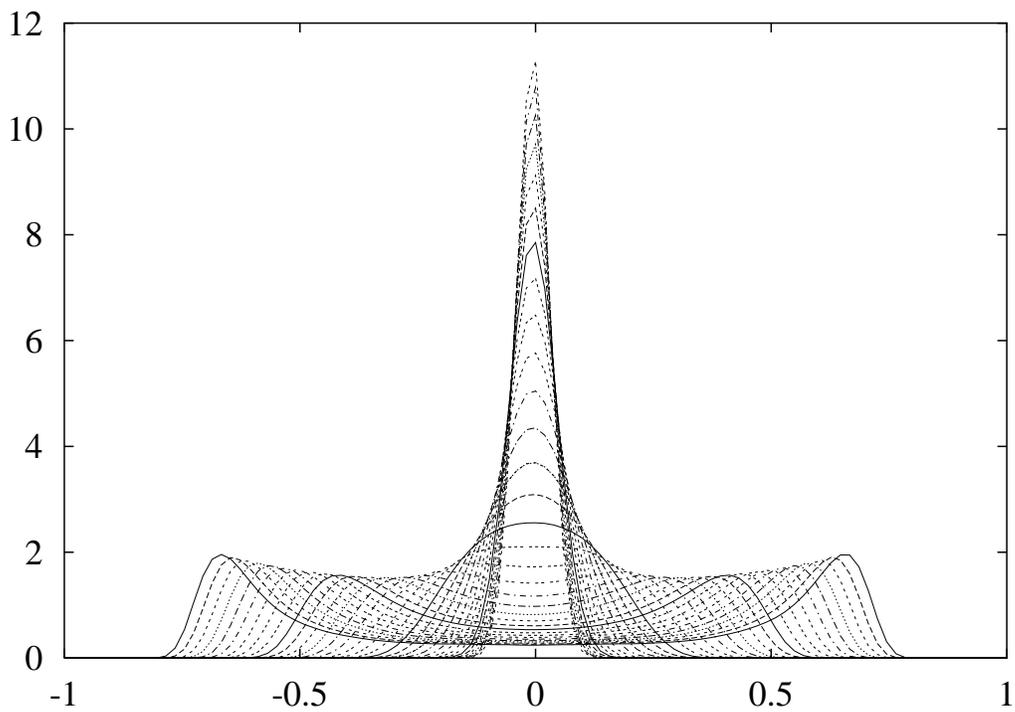}
\caption[qqq]{$P(q)$ at $L=8$, 
for all values of $T$. From single peak in $q=0$ to continuous part and
double peak at large $q$ for increasing $T$.}
\label{fig:pqL8}
\end{figure}

In figure \figref{fig:pqL8} we compare $P(q)$ at different values of
$T$ on the same lattice size. From the single peaked shape at high $T$
one gets a clear double peaked structure, with a clear {\em plateau}
at low $q$, in the low $T$ region.

\begin{figure}
\centering
	\includegraphics[width=0.8\textwidth]{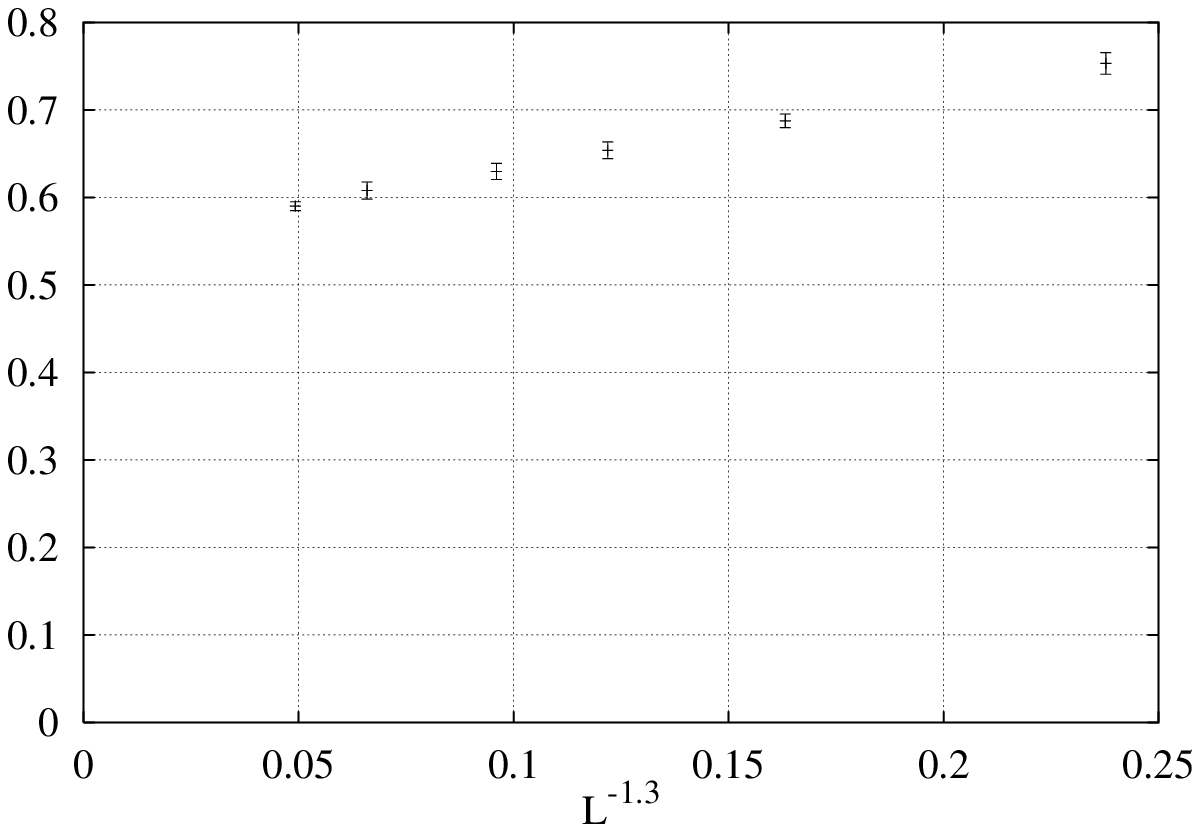}
\caption[qqq]{$q_{max}$ versus $L^{-1.3}$.}
\label{fig:qmL13}
\end{figure}

As we have said, in order to establish that we are having a real phase
transition in the infinite volume limit we have to show that the value
of $q=q_{max}$ where $P(q)$ is maximum does not go to zero. We start
by plotting in figure \figref{fig:qmL13} $q_{max}$ versus $L^{-1.3}$
(the exponent $1.3$ comes from our best fit, see later). It is easy to
see that an asymptotic value $q_{max}=0$ seems unplausible.

\begin{figure}
\centering
	\includegraphics[width=0.8\textwidth]{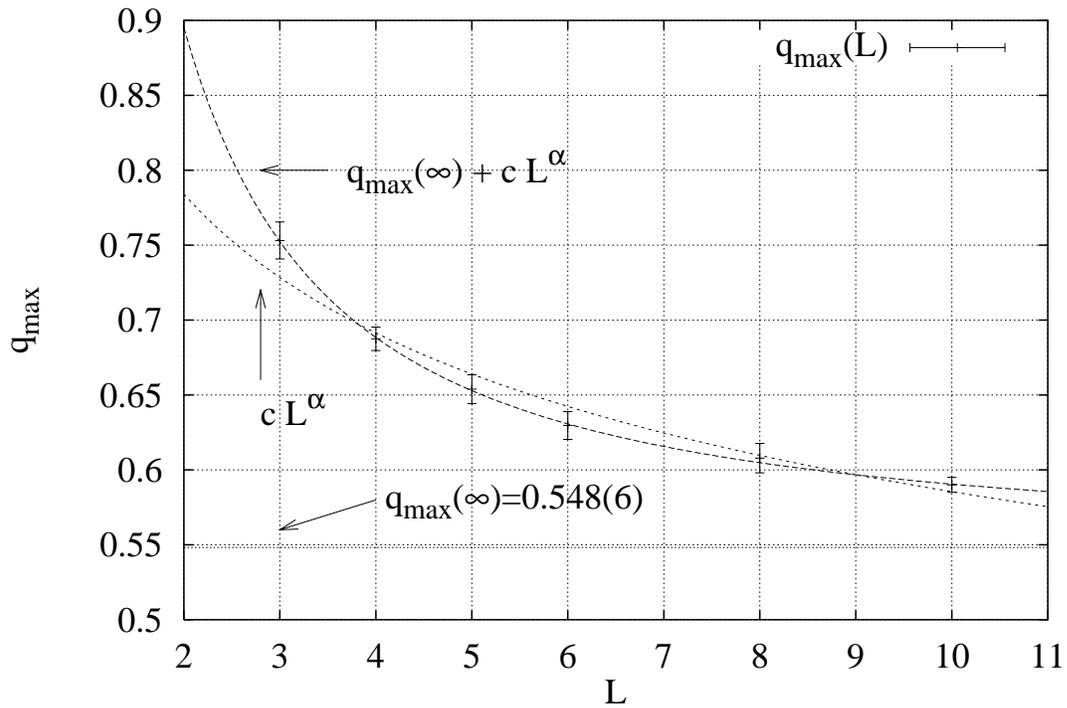}
\caption[qqq]{$q_{max}$ versus $L$ and our two best fits.}
\label{fig:qmax_t1300}
\end{figure}

Making this last statement more quantitative needs a more careful
analysis. In order to do that we fit

\begin{equation}
  q_{max}(L) =   q_{max}(\infty) + \frac{A}{L^{\alpha}}\ , 
\end{equation}
both with $q_{max}(\infty)=0$ and by allowing for it a non zero
value. In figure \figref{fig:qmax_t1300} we plot the values of
$q_{max}$ versus $L$ and the results of the two best fits, one with a
fitted value of $q_{max}(\infty)$ and the second with fixed  
$q_{max}(\infty)=0$. This second fit is clearly unsuitable, and it has
a very high value of $\chi^2$. In the first fit we get

\begin{equation}
  q_{max}(\infty) = 0.548 \pm 0.006\ ,
\end{equation}
that is our best estimate for the position of the $\delta$ function at
$q_{EA}$ in the infinite volume limit. We estimate $\alpha=1.3\pm 0.1$
(in the fit with a zero asymptotic value one finds the very small
value $\alpha\simeq 0.2$).

\begin{figure}
\centering
	\includegraphics[width=0.8\textwidth]{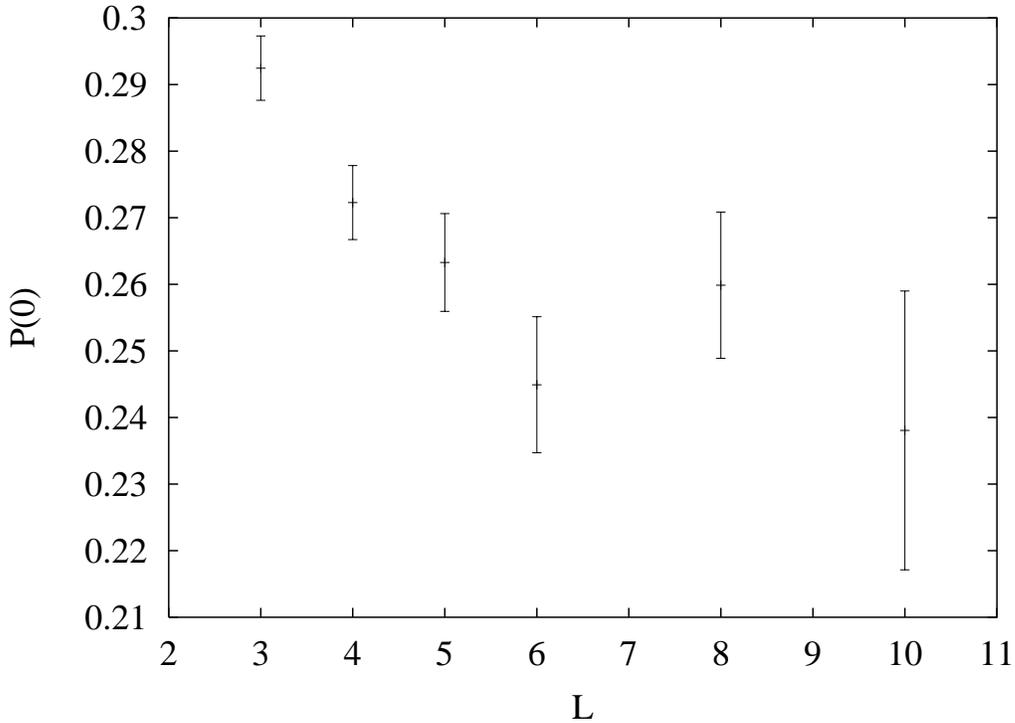}
\caption[qqq]{$P(q\simeq 0)$ versus $L$. $T=1.2$.}
\label{fig:p0}
\end{figure}

In figure \figref{fig:p0} we show the value of $P(q)$ close to $q=0$
(averaged over a small $q$ range, where $P(q)$ is remarkably constant,
in order to diminish statistical fluctuations) as a function of
$L$. One cannot observe any statistically significant decrease of this
value for increasing large lattice volume (there is a small decrease
only for small volumes).  The most plausible implication of this
evidence is that the system has many stable states, and that the cold
$T$ phase is characterized by Replica Symmetry Breaking (even if it
has to be stressed that this evidence is not as strong as the one
implied by the figure \figref{fig:qmax_t1300}).

\begin{figure}
\centering
	\includegraphics[width=0.8\textwidth]{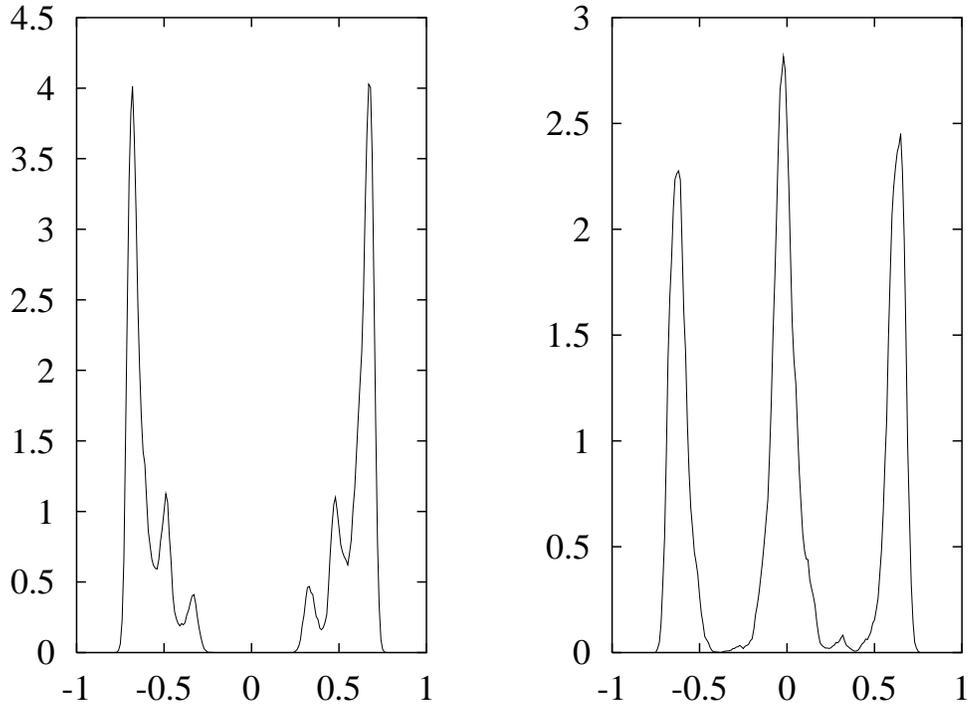}
\caption[qqq]{$P_J(q)$ for selected samples. $T=1.2$, $L=10$.}
\label{fig:pjq}
\end{figure}


\begin{figure}
\centering
	\includegraphics[width=0.8\textwidth]{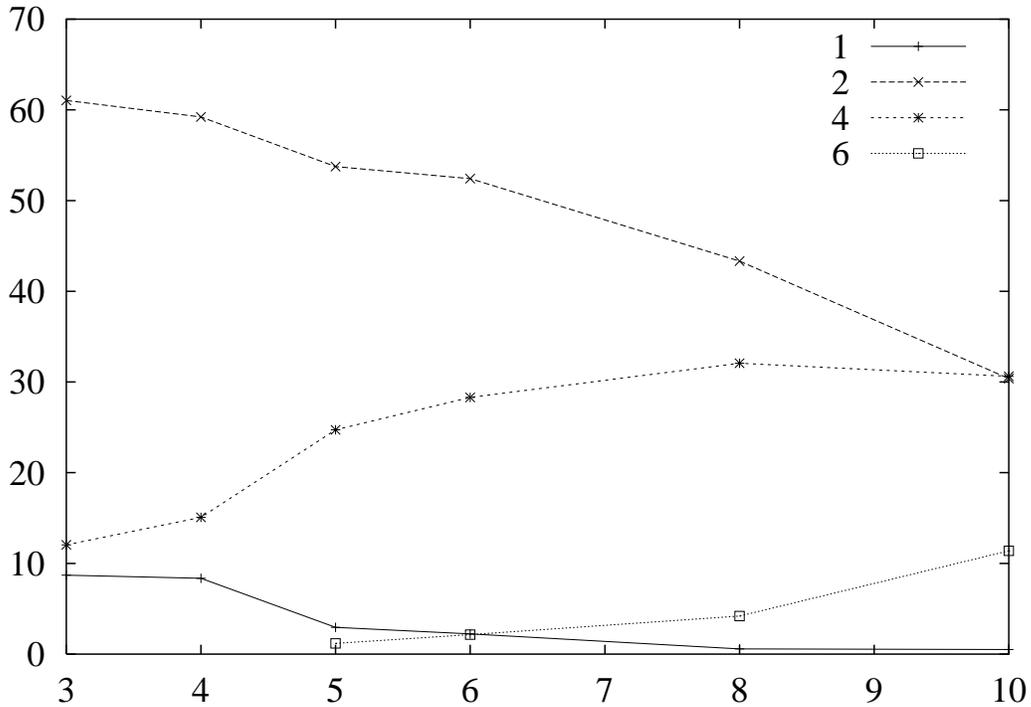}
\caption[qqq]{
Percentage of disorder
configurations such that $P_J(q)$ has 
$1$, $2$, $4$ and $6$
peaks versus $L$. The number of configurations with a complex phase
space ($P_J(q)$ with many peaks) increases strongly with $L$.}
\label{fig:isto_max}
\end{figure}

In figure \figref{fig:pjq} we plot $P_J(q)$ for selected samples, at
$T=1.2$, $L=10$. One can see here that they are very complex
distributions: such a pattern is typically related to the presence of
many states (it has to be notice however that the small side peaks are
not always there because of the presence of a real state).

To be more quantitative we have measured the percentage of disorder
configurations such that $P_J(q)$ has $1$, $2$, $4$ and $6$ peaks
versus $L$, and we plot it in figure \figref{fig:isto_max}. The
number of configurations with a complex phase space ($P_J(q)$ with
many peaks) increases strongly with $L$. We use this evidence to rule
out the picture of a {\em modified droplet model}, that has been
discussed, among others, in \cite{NS} and in references therein. The
picture of the modified droplet models implies that for each
realization of the quenched disorder there are (in the cold phase)
only two ground states, but that the value of $q_{EA}$ (i.e. the
support of the $\delta$ function that constitutes the $P_J(q)$)
depends on the sample. Here, on the contrary, the number of states for
a given sample is strongly increasing with $L$ (and with decreasing
$T$).

\section{Sum Rules}

\begin{figure}
\centering
	\includegraphics[width=0.8\textwidth]{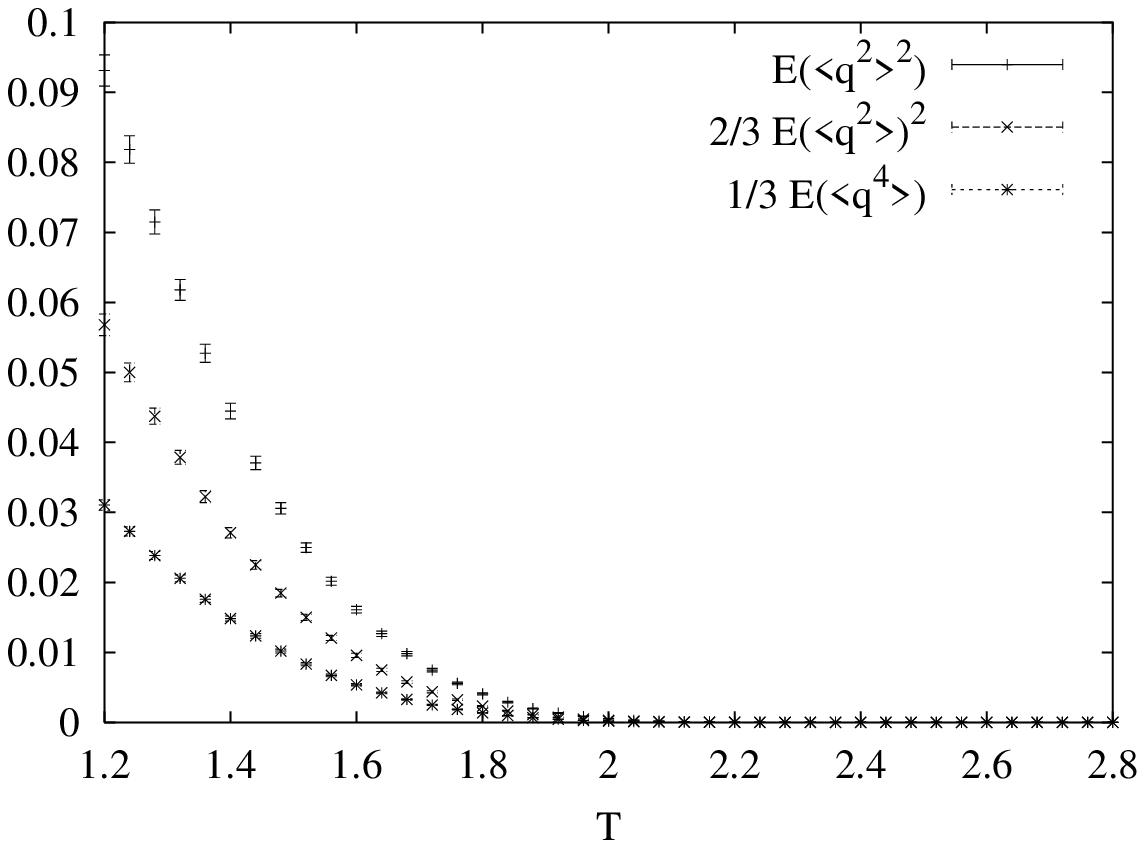}
\caption[qqq]{
  $\md{\mt{q^2}^2}$, $\md{\mt{q^2}}^2$ 
  and $\md{\mt{q^4}}$ vs. $T$.
}
\label{fig:lhs_rhs_sumrule}
\end{figure}

In this section we discuss another important feature of the broken
phase of the $4D$ EA model.  The starting point for this analysis can
be for example the relation:

\begin{equation}
               \md{\mt{q^2}^2}    =
          \frac23  \md{\mt{q^2}}^2 +
          \frac13  \md{\mt{q^4}} \ .
\protect\label{G-A}
\end{equation}
This is one of a set of relations that are valid in the Mean Field
Theory of Spin Glasses \cite{MPSTV}.  The work contained in
\cite{SUMRULE} has established numerically that these relations are
satisfied with good accuracy also in finite dimensional spin
glasses. Following these findings a rigorous and theoretical analysis
has improved our understanding of such set of sum rules
\cite{GUERRA,AIZCON,PARISISR,NSLONG}: they are strongly related to the
ultrametric properties of the phase space.

First of all we show evidence that the relation (\ref{G-A}) has a
non-trivial content in the low-temperature phase (in the high $T$
phase it is satisfied in the form $0=0$).  Figure
\figref{fig:lhs_rhs_sumrule} shows that the values of the three
quantities involved in (\ref{G-A}) are significantly different from
zero below $T_c$ (we have already shown in better detail that the
infinite volume of such quantities is non-zero).

\begin{figure}
\centering
	\includegraphics[width=0.8\textwidth]{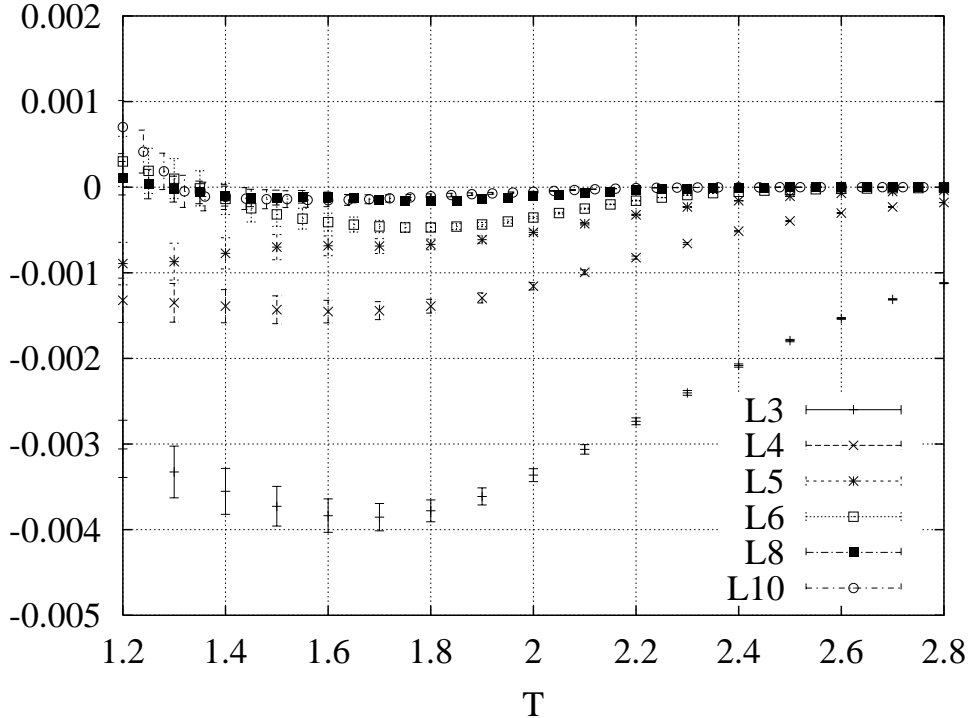}
\caption[qqq]{Left hand side minus right hand 
side of equation \eref{G-A} vs. $T$.}
\label{fig:lhs_minus_rhs_sumrule}
\end{figure}

In figure \figref{fig:lhs_minus_rhs_sumrule} we show the difference
between the left hand side and the right hand side of \eref{G-A}.  The
two contributions cancel out with good accuracy (to $3$ significant
figures), and asymptotically for large lattice size the difference
extrapolates to zero.

\begin{figure}
\centering
	\includegraphics[width=0.8\textwidth]{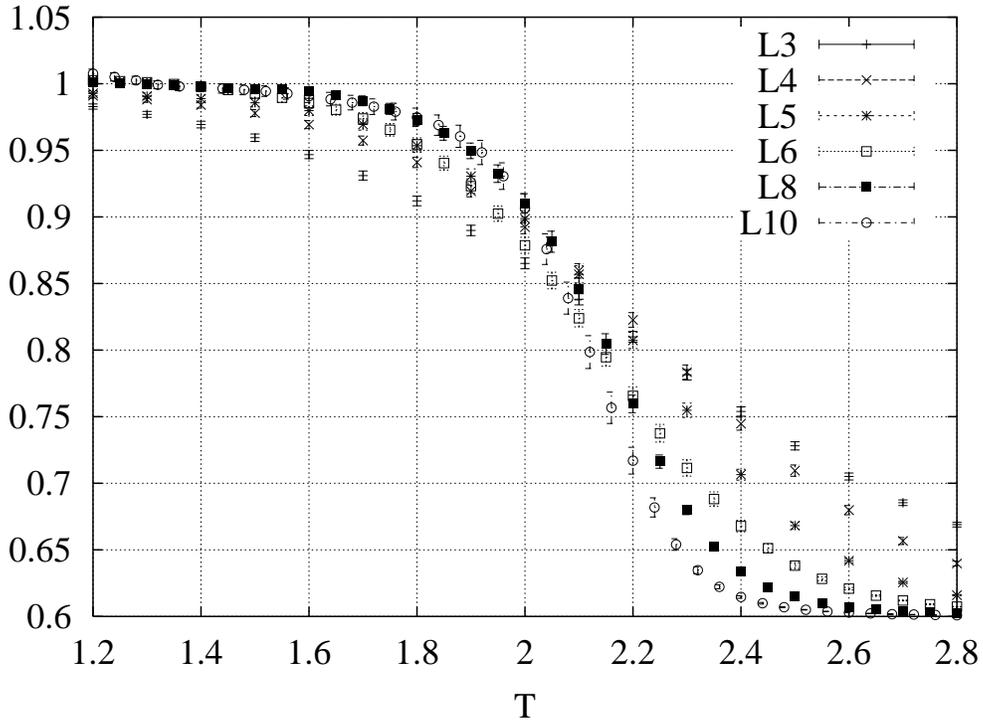}
\caption[qqq]{Ratio of left hand side and right hand side 
of equation \eref{G-A} versus $T$.}
\label{fig:ratio_sumrule}
\end{figure}

Another possible way to visualize the result is plotting the ratio of
the left hand side and the right hand side.  As figure
\figref{fig:ratio_sumrule} shows, for $T$ below $T_c$ we get
identically one, while for $T \to \infty$ we get the value $\frac35$,
expected for a Gaussian $P(q)$.

\begin{figure}
\centering
	\includegraphics[width=0.8\textwidth]{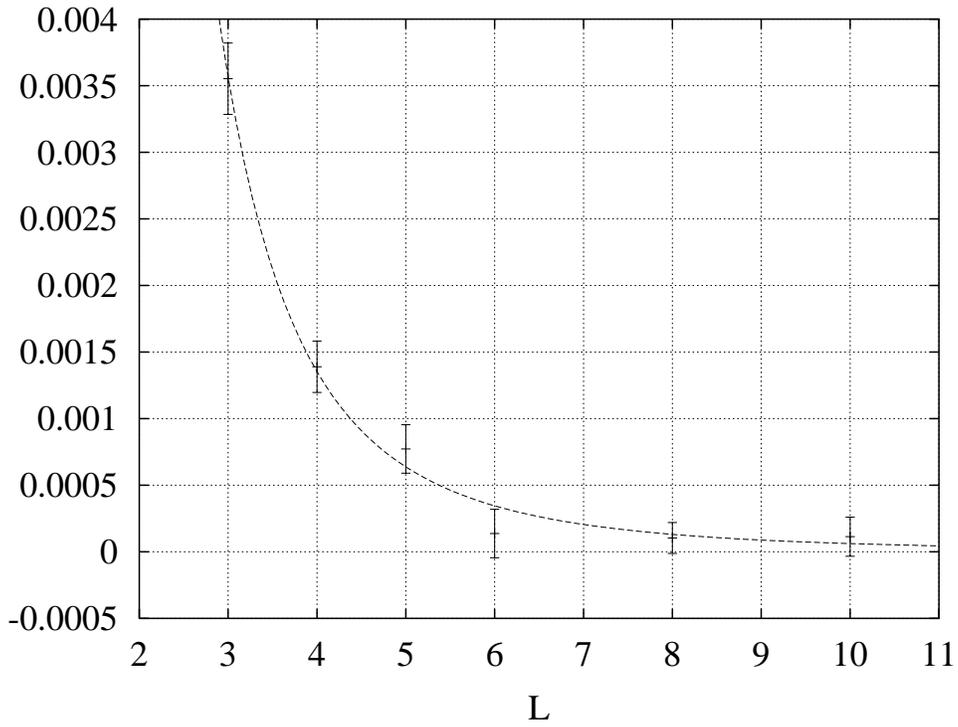}
\caption[qqq]{
Left hand side minus right hand 
side of equation \eref{G-A}
versus $L$, at $T=1.4$, and 
and best fit to a zero constant value with 
a simple power correction.}
\label{fig:scaling_sumrule}
\end{figure}

We have also fitted the difference plotted in figure
\figref{fig:lhs_minus_rhs_sumrule}, for various values of temperatures
$T<T_c$: in all cases a fit to an asymptotic zero value with power
corrections works very well, and the exponent of the corrections is
close to $3$ for all $T$ values. As an example we plot the data
together with the best fit for $T=1.4$ in figure
\figref{fig:scaling_sumrule}.

\section{Conclusions}

In this note we have been able to give strong evidence for mean field
behavior of the $4d$ Ising spin glass with binary couplings. Life in
the $4d$ model is easier than in $3d$, where even after a large number
of intense numerical simulations the evidence for a phase transition
is still slightly marginal (even if, at this point, convincing
enough). In our case already the crossing of the Binder cumulants is
the very clear signature of a typical phase transition (as opposed to
the quasi-merging, quasi-Kosterlitz-Thouless behavior of the $3d$
case). It is clear that $3d$ is very close to the lower critical
dimension, and that there observing the effects of the physical
critical point is dramatically difficult. $4d$ is on the safer side,
and numerical simulations show that very clearly.

We have been able to determine critical exponents precisely, and
to enter in the large volume region with good accuracy. For example
we have been able to show that the peak of $P(q)$ is not going to
$q=0$ for increasing lattice size, and (with a slightly worst accuracy
and level of reliability) that the plateau at $q\simeq 0$ does not
decrease with increasing lattice size. Also we remind the reader that
non-trivial sum rules are satisfied with very good accuracy. So,
thinks look quite clear in the $4d$ case.

\section*{Acknowledgments}

We thank Giorgio Parisi, Federico Ricci-Tersenghi and
Juan Ruiz-Lorenzo for a number of interesting discussions.

\end{document}